\documentclass[aps,prl,twocolumn,epsfig,amsmath,floatfix,,showpacs]{revtex4}

\usepackage{epsfig,amsmath}
\usepackage{graphicx}
\usepackage{dcolumn}
\usepackage{bm}

\newcommand{\beq}{\begin{equation}}
\newcommand{\eeq}{\end{equation}}
\newcommand{\beqa}{\begin{eqnarray}}
\newcommand{\eeqa}{\end{eqnarray}}
\newcommand{\lam}{\lambda} 
 
\newcommand{\la}{\langle} \newcommand{\cP}{{\cal P}}
\newcommand{\ra}{\rangle} \newcommand{\cC}{{\cal C}}

\newcommand{\non}{\nonumber} 
\newcommand{\rp}{r_\perp} 
\newcommand{\rDe}{|{\bf e}\ra} \newcommand{\rDE}{|{\bf E}\ra}
\newcommand{\lDe}{\la{\bf e}}

\def\cp#1{{ Contem.\ Phys.} {\bf#1}}

\def\natphot#1{{ Nat.\ Phot.} {\bf#1}}
\def\njp#1{{ New\ J.\ Phys.\/} {\bf#1}}
\def\njp#1{{ New\ J.\ Phys.\/} {\bf#1}}

\def\oc#1{{ Opt.\ Commun.} {\bf#1}}

\def\pra#1{{ Phys.\ Rev. A\/} {\bf#1}}

\def\prl#1{{ Phys.\ Rev.\ Lett.} {\bf#1}}
\def\pscr#1{{ Phys.\ Scrip.} {\bf#1}}

\begin{document}

\title{Coherence Constraints and the Last Hidden Optical Coherence}
\author{Xiao-Feng Qian$^{1,3}$}
\email{xfqian@pas.rochester.edu}
\author{Tanya Malhotra$^{1,2}$}
\author{A. Nick Vamivakas$^{1,3}$}
\author{Joseph H. Eberly$^{1,2,3}$}
\affiliation{$^{1}$Center for Coherence and Quantum Optics,
University of Rochester,
Rochester, New York 14627, USA\\
$^{2}$Department of Physics \& Astronomy, University of Rochester,
Rochester, New York 14627, USA\\
$^{3}$The Institute of Optics, University of Rochester, Rochester, NY 14627,
USA}

\date{\today }

\begin{abstract}
We have discovered a new domain of optical coherence, and show that it is the third and last member of a previously unreported fundamental triad of coherences. These are unified by our derivation of a parallel triad of coherence constraints that take the form of complementarity relations. We have been able to enter this new coherence domain experimentally and we describe the novel tomographic approach devised for that purpose.

\end{abstract}


\maketitle


\noindent{\bf Background:} Coherence is a concept whose entrance into physics can be traced to Young's report of light field interference \cite{Young-02}. Its importance is now recognized across all of science from astronomy to chemistry and biology as well as  in physics (see \cite{Glauber, Mandel-Wolf-95}). This is the reason that recent announcements of previously hidden optical coherencesÊ\cite{Kagalwala-etal-13, DeZela-14, Eberly-15, Svozilik-etal-15, Khoury-etal-15b, Eberly-etal-16} have been so startling. In our view, a unified explanation has been strikingly absent. Here we report discovery of a new domain of optical coherence. We also show that it is the third and last member of a fundamental triad of coherences, and additionally that these are unified by the derivation of a triad of previously unreported coherence constraints that take the form of complementarity relations. They provide a coordinated understanding of all so-called hidden coherences. Accompanying these advances is a report of experimental entry into the new coherence domain, and details of results obtained within it.Ê

We begin by accounting for the independent degrees of freedom available to an observed optical field. These are space, time, and spin (intrinsic polarization). The idealized optical beam context, with a single direction of propagation, allows a slight simplification, which we take for granted. We will ignore the propagation degree of freedom and write the beam's complex amplitude in terms of the orthonormal bases for each of its other degrees of freedom, the two-dimensional transverse coordinate $\rp$, time $t$ and spin (polarization) $\hat s$:
\beq \label{Efield}
\vec E(\rp, t) = E_0  \sum_{k,m}\sum_{i=1,2}d_{ikm}\hat s_i F_k(t) G_m(\rp),
\eeq
where $d_{ikm}$ are complex coefficients. Specifically, the spin unit vectors $\hat s_i$, conventionally ($\hat h$, $\hat v$)   or ($\hat x$, $\hat y$), satisfy $\hat s_1 \cdot \hat s_2 = 0$, and the transverse beam  basis functions $G_m(\rp)$ are taken as orthonormal in integration across the beam. The basis functions $F_k(t)$ are orthonormal eigenfunctions of the integral equation that has the field's temporal correlation function as kernel (see \cite{Kac-Siegert-47} and Sec. 4.7.1 in \cite{Mandel-Wolf-95}). 

The foundation of our analysis is a finite set of contextual  projections of $\vec E$. They are accomplished by more or less obvious experimental arrangements, and their purpose is to isolate independent coherences among pairs of degrees of freedom. Each of these degrees of freedom defines (occupies) one of the independent vector spaces of the field, and for convenience we now label them $s$\ for spin, $t$\ for time, and $r$\ for transverse spatial location. For example, the $r$ projection is accomplished by transverse beam integration, leaving an $st$ field:
\beqa \label{vecEst}
\vec E^{(m)}_{st}(t) &=& \int d^2\rp G^*_m(\rp)\vec E(\rp, t) \non \\
&=& E_0 \sum_{k} \sum_{i=1,2}d_{ikm} \hat s_i F_k(t).
\eeqa
Each of the $s,\ t,\ r$\ vector spaces allows such a projection, and the resulting projections of the field (\ref{Efield}) produce these three reduced vectors:
\beqa \label{efield}
\rDe_{tr} &=& \sum_{k,m}a_{km} |F_k\ra \otimes |G_m\ra, \label{etr}\\
\rDe_{sr} &=& \sum_{i,m}b_{im}|s_i\ra \otimes |G_m\ra, \label{esr}\\
\rDe_{st} &=& \sum_{i,k}c_{ik}|s_i\ra \otimes |F_k\ra, \label{est}
\eeqa
where the tensor product symbols between vector spaces will rarely be repeated. The three relations (\ref{etr})-(\ref{est}) all refer to the same original field (\ref{Efield}). They arise in three different experimental contexts \cite{moredegrees}. The lower case $\rDe$ notation indicates that each of the projected fields has been normalized to unit intensity. The coefficients contain relative amplitudes making the kets orthonormal in their own vector spaces: $\la s_i|s_j\ra = \la F_i|F_j\ra = \la G_i|G_j\ra = \delta_{ij}$, and $\lDe\rDe = 1$ in each case. 

To discuss the coherence in (\ref{est}), which is the most familiar projection and the same as in (\ref{vecEst}), we can conveniently denote the orthogonal components $|s_i\ra$ in (\ref{est}) to be horizontal and vertical: $|h\ra$ and $|v\ra$. Then we have
\beq \label{Dexy}
\rDe_{st} = \sum_{k} c_{hk}|h\ra| \otimes |F_k\ra  + \sum_{k}c_{vk}|v\ra| \otimes |F_k\ra.\\
\eeq
The polarization (spin) vectors $|h\ra$ and $|v\ra$ identify their respective temporal mode sums as the independent horizontal and vertical components of the normalized field:
\beq
\sum_k c_{hk}|F_k\ra = |e_h\ra \quad {\rm and}\quad \sum_k c_{vk}|F_k\ra = |e_v\ra,
\eeq
allowing us to rewrite (\ref{est}) and (\ref{Dexy}) as:
\beq \label{FxFy}
\rDe_{st} = \cos\frac{\theta}{2} |h\ra \otimes |e_h\ra + \sin\frac{\theta}{2}|v\ra \otimes |e_v\ra,  
\eeq
where the cosine and sine factors permit arbitrary division of the unit amplitude between the two terms while allowing each component to be unit-normalized: $\la e_h|e_h\ra = \la e_v|e_v\ra =1$. 

Equation (\ref{FxFy}) shows that the spin and amplitude vector spaces making up $\rDe_{st}$ are factorable when $\rDe_{st}$ is perfectly polarized. For example, if $\theta = \pi$, then the field is completely $v$ polarized (has only a $|v\ra$ component). By the same token $\rDe_{st} \to |v\ra \otimes |e_v\ra$ gives a field that is obviously factorable (separable) between its spin and temporal-amplitude degrees of freedom.\\ 

\noindent{\bf Coherence Constraints and a New Coherence Domain:} \quad The noted similarity of polarizability and separability can be quantified. The degree of polarization $\cP_{st}$ is determined by the two eigenvalues $\lam_1 \ge \lam_2$ of the polarization coherence matrix, and they obey $\lam_1 + \lam_2 = 1$ when the field is normalized to unit intensity, as we have done, giving the known result  \cite{Wolf-59, Brosseau}
\beq \label{cP}
\cP_{st} = \lam_1 - \lam_2,
\eeq
guaranteeing $1 \ge \cP_{st} \ge 0$. At the same time, the degree of $st$ entanglement, which we measure via concurrence \cite{Wootters} and denote $\cC_{st}$, is given by 
\beq \label{cCst}
\cC_{st} = 2\sqrt{\lam_1\lam_2},
\eeq
where $1 \ge \cC_{st} \ge 0$. Simple arithmetic now yields a  quadratic constraint. It quantifies the coherence-sharing which unites degree of polarization and degree of concurrence (non-separability, entanglement):
\beq \label{CPst}
\cC^2_{st} + \cP^2_{st} = 1.
\eeq

This constraint is significant, not coincidental \cite{Fedorov-etal-11}. Its exact counterpart arises in the independent (and contextually distinct) $sr$ coherence present in (\ref{esr}). A form of $sr$ coherence was perhaps first noted by Gori, et al. \cite{Gori-etal-06}, and has recently been explored in detail both experimentally and theoretically and noted as a hidden coherence by Abouraddy, et al. \cite{Kagalwala-etal-13}. They employed the Bell measure to engage entanglement. The joint {\it sr} correlation also supports a polarization matrix, independent of the $st$ correlation but with the same eigenvalue properties, so the same constraint also applies to $sr$ coherence:
\beq \label{CPsr}
\cC^2_{sr} + \cP^2_{sr} = 1.
\eeq

Given the three projection relations (\ref{etr})-(\ref{est}), it is obvious that there must exist a $tr$ coherence. In a striking departure from previous cases, $tr$ coherence implies a new kind of ``polarization", one in which the spin degree of freedom is not involved at all. This opens a door on an unexplored domain of optical coherence, the heretofore missing member of a fundamental triad implied by the triad of degrees of freedom of the field. Clearly it must be included for completeness \cite{moredegrees}. 

In the following Sections we will report the first experimental observations and quantifications associated with it, as well as a laboratory search for the now-expected third quadratic constraint:
\beq \label{CPtr}
\cC^2_{tr} + \cP^2_{tr} = 1.
\eeq
 
The doubly infinite summation of the temporal and spatial modes in (\ref{etr}) means that the $tr$ coherence matrix is infinite-dimensional in the most general case, with an infinite set of non-zero eigenvalues. This precludes the use of the same analysis of polarization and entanglement employed for $st$ and $sr$ coherences. However, there is an open route of eigenvalue analysis via the Schmidt Theorem of analytic function theory \cite{Schmidt}, which we have demonstrated previously \cite{Qian-Eberly-11}, but we need less than that here. As we have insisted, coherence is highly contextual, and the appropriate $tr$ coherence for initial attention is the one arising from a $tr$ context that allows comparison with $st$ and $sr$ coherences. This is done in the next Section with an experimental setup that enters the new coherence domain via two orthonormal spatial modes.\\

\noindent{\bf Experimental Considerations:} \quad Our experimental $tr$ analysis uses an optical field running in two Hermite-Gauss (HG) orthonormal spatial modes, i.e., HG\{10\} and HG\{01\} that we designate $|G_a\ra$ and $|G_b\ra$ respectively. After projection on an arbitrary direction, say horizontal, of spin-polarization, e.g., as in the projection (\ref{etr}), the field (not intensity-normalized) can be written as
\beq \label{xDEtr}
\la h\rDE = \rDE_{tr} = |E_a\ra \otimes |G_a\ra + |E_b\ra \otimes |G_b\ra, 
\eeq
where the two amplitudes $|E_a\ra$ and $|E_b\ra$ represent combinations of many orthonormal temporal modes $|F_k\ra$:
\beq \label{xDEtr}
|E_a\ra = \sum_k a_{ka}|F_k\ra,\ {\rm and}\ |E_b\ra = \sum_k a_{kb}|F_k\ra.
\eeq
They comprise an unknown combination of modes that we take experimentally from a multimode diode laser running below threshold. Because $\la G_a|G_b\ra = \delta_{ab}$, the intensity is given by $I = |\la h\rDE|^2 = \la E_a|E_a\ra + \la E_b|E_b\ra$. 

For comparison with its $st$ counterpart in (\ref{FxFy}), we can unit-normalize this $tr$ field. We can again use an unspecified angle $\theta$ to signal a division of intensity between $I_a$ and $I_b$, meaning a division of field strength between $|E_a\ra$ and $|E_b\ra$. Thus we express the non-orthogonal $|E\ra$s in terms of also non-orthogonal but unit-normalized $|e\ra$s to get
\beq
|E_a\ra = \sqrt{I}\cos\frac{\theta}{2}|e_a\ra, \quad\ |E_b\ra = \sqrt{I}\sin\frac{\theta}{2}|e_b\ra. 
\eeq
By including the two $|G\ra$ modes and removing the $\sqrt{I}$ factors we obtain the exact analog of (\ref{FxFy}), with the unit-normalized $|G_a\ra$ and $|G_b\ra$ replacing the $|h\ra$ and $|v\ra$ unit vectors. Then the unit-normalized field is written
\beq \label{Detr}
\rDe_{tr} = \cos\frac{\theta}{2} |G_a\ra \otimes |e_a\ra + \sin\frac{\theta}{2}|G_b\ra \otimes |e_b\ra  , 
\eeq
where $\la e_a|e_a \ra = \la e_b|e_b \ra = 1$, and
\beq
\la e_a|e_b \ra \equiv \gamma \equiv \delta e^{i\phi}.
\eeq

Thus the term by term alignment of (\ref{FxFy}) and (\ref{Detr}) makes it clear that the earlier constraints, (\ref{CPst}) and (\ref{CPsr}), should have an exact counterpart here: $\cC^2_{tr} + \cP^2_{tr} = 1$. This conclusion should be examined and tested, and we do that. See the last column of Table \ref{StokesTable}.

The Stokes vector analogs that we have recorded for this new coherence are defined in the standard way (see an early consideration by Padgett and Courtial in \cite{Padgett-Courtial-99}):
\beqa  \label{S0S1S2S3}
S_0 &=& I_a + I_b = I \\
S_1 &=& I_a - I_b = I\cos\theta \\
S_2 &=& \la F_a|F_b\ra + \la F_b|F_a\ra \non \\
&=& I\ \delta\sin\theta\cos\phi\\
S_3 &=& i[\la F_a|F_b\ra - \la F_b|F_a\ra] \non \\
&=& I\ \delta\sin\theta\sin\phi,
\eeqa
where $\delta$, $\theta$ and $\phi$ are defined above. The radius of the sphere normalized to $S_0$ is conventionally called the degree of polarization, which refers here to time-space coherence, so we can write 
\beq \label{radius} 
{\cal P}^2_{tr} = \frac{S_1^2 + S_2^2 + S_3^2}{S_0^2} =  \cos^2\theta + \delta^2\sin^2\theta. 
\eeq 
Clearly $\theta$ and $\delta$ control the radius of the Poincar\'e sphere and provide total (unit radius) time-space coherence  when either $\delta = 1$ or $\theta = \pi$, and a reduced sphere radius implying only partial coherence otherwise.\\

\begin{figure}[t!]
\includegraphics[width=7.cm]{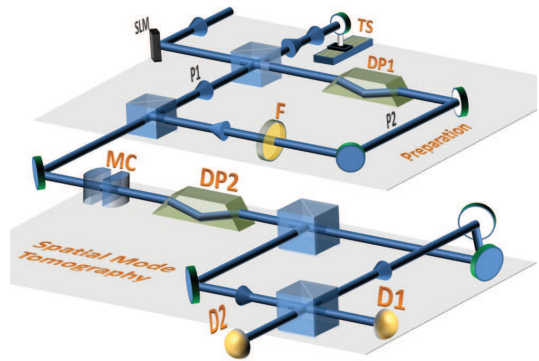}
\caption{Schematic experimental setup. Light field (\ref{xDEtr}) is prepared with a spatial light modulator (SLM) and a modified Mach-Zehnder interferometer (MZI), where filter (F) controls $\theta$ and the translation stage (TS) manages $\delta$ and $\phi$. The Stokes parameters are measured with different combinations of a mode converter (MC), a Dove prism (DP2), and a Mach-Zehnder interferometer with an additional Mirror (MZIM).} 
\label{setup}
\end{figure}

\noindent\textbf{Spatial Mode Coherence Tomography:} \quad We have implemented a new experimental tomography procedure that is able to acquire complete information of an arbitrary unknown two-mode input state made from $G_{a}$ and $G_{b}$. An arbitrary $tr$\ optical beam of the form in (\ref{xDEtr}) is a good example. The experimental setup for this novel {\it tr}\ tomography is illustrated by Fig.~\ref{setup}. In the preparation stage, a spatial light modulator (SLM) is used to generate a specific transverse mode, i.e., $ E_{in}(r_\perp,t) =G_{a}(r_\perp)F_{a}(t)$. It is then sent through a Mach-Zehnder interferometer (MZI) with two ordinary 50/50 beam splitters. In Path 1 (P1), the statistical temporal amplitude $F_{a}(t)$ is delayed with a translation stage (TS), while in Path 2 (P2), a Dove prism (DP1) oriented at $\pi /4$ is used to rotate the spatial mode $G_{a}$ into $G_{b}$ and a filter (F) is placed to adjust the path intensity. The output beam of the MZI is in exactly the form $E(r_\perp,t) = G_{a}(r_\perp)F_{a}(t) + G_{b}(r_\perp)F_{b}(t)$, of which the normalized expression is given by Eq.~(\ref{Detr}). The coefficients $\cos (\theta/2)$ and $\sin(\theta/2)$ are controlled by the filter in P2 and the parameters $\delta$ and $\phi$ are managed by adjusting the delay in P1.

The {\it tr} coherence tomography stage is composed of three major elements, a spatial mode converter (MC), a Dove prism (DP2), and a Mach-Zehnder interferometer with an additional mirror (MZIM) \cite{MZIM}. These elements are respectively exact analogs of a quarter-wave plate, half-wave plate, and polarizing beam splitter that are employed in ordinary {\it st} or {\it sr} spin-polarization tomography. The mode converter MC contains a pair of appropriately separated cylindrical lenses that will introduce a relative $i$ phase to the $G_{b}$ mode with respect to $G_{a}$ \cite{MC}. The Dove prism is used to rotate the spatial modes $G_{a}$ and $G_{b}$ to a desired basis $\alpha G_{a}+\beta G_{b}$. The MZIM is employed to project the mode states $G_{a}$, $G_{b}$ onto the two output ports respectively.

\begin{figure}[h!]
\includegraphics[width=5cm]{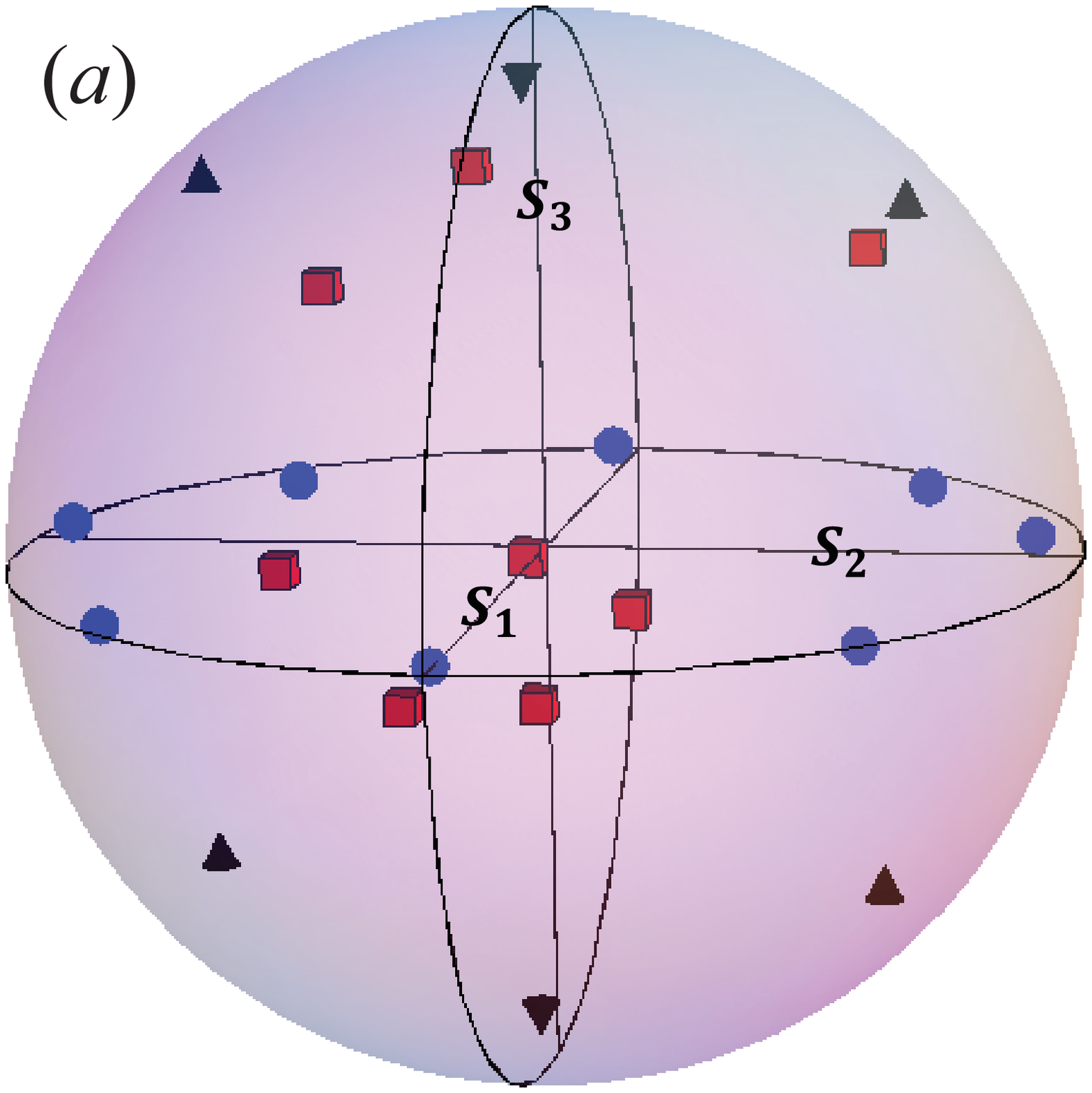}
\includegraphics[width=4cm]{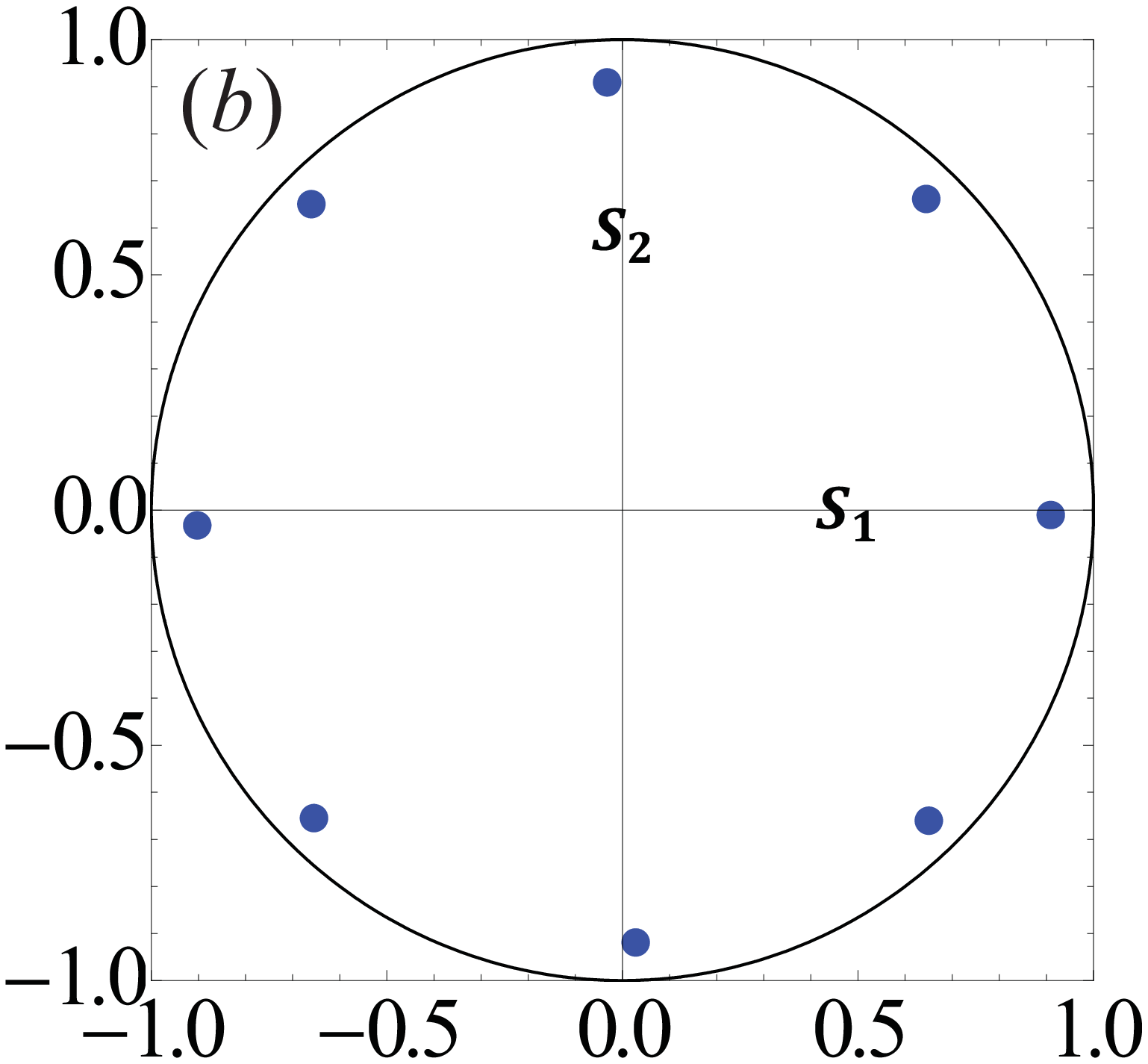}
\includegraphics[width=4cm]{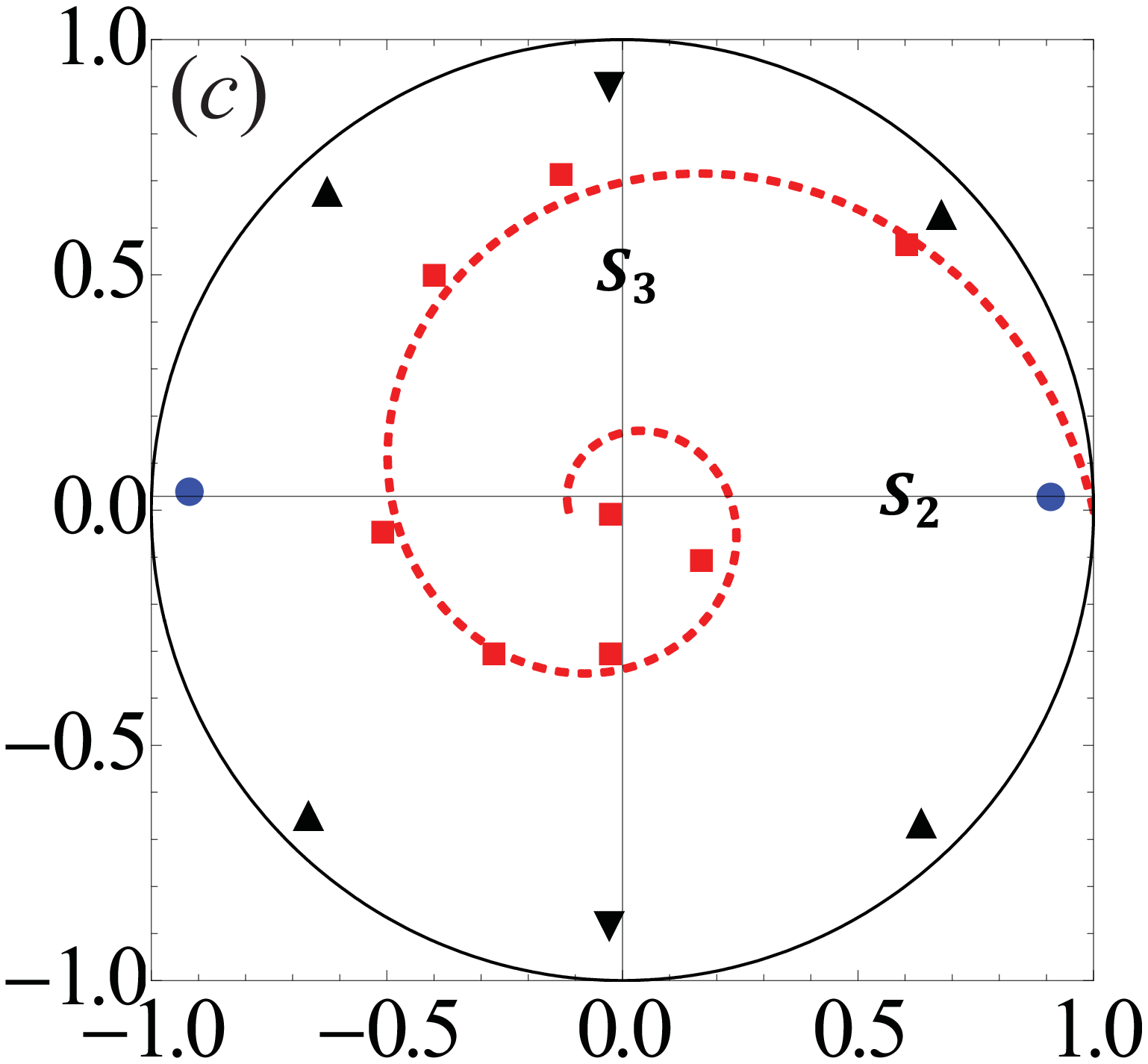}
\caption{Experimental data of various time-space polarization states. Plot (a) is a Poincar\'{e} sphere representation of all measured states where the blue dots, black triangles and red squares denote linear, elliptical (including circular) and partial (including completely un-polarized) polarization states respectively. Plot (b) shows points on the equatorial plane $S_3=0$, and it contains all the linear $tr$ states. Plot (c) shows points in the $S_1=0$ plane where the black upward triangles, black downward triangles, and red squares denote elliptical, circular and partially coherent $tr$ states, respectively. The red dotted curve is the logarithmic polar spiral function $\delta = e^{-0.23\phi}$ tracking states with smaller and smaller degrees of {\it tr} coherence. Note: the error bars of the measured Stokes parameters are relatively small and not shown, but the magnitude of the maximum error is given in Table I.} \label{results}
\end{figure}


With the combination of the MZIM and the Dove prism DP2 (oriented appropriately in the rotated basis, $G_{a}$, $G_{b}$ and $G_{a} \pm G_{b}$), we are able to obtain the Stokes parameters $S_{0}$, $S_{1}$, and $S_{2}$. Accordingly, the combination of all three elements, with the DP2 and MC adjusted to account for a rotated basis $G_{a} \pm i G_{b}$, amounts to an effective measurement of $S_{3}$. Therefore all four Stokes parameters can be recorded. \\

The conventional presentation of spatial modes for an optical beam is to display irradiance images of the transverse plane. These provide a positive visual validation of mode quality, but with the experimental setup described we can do much more. We produced and reconstructed the states of various {\it tr} ``polarized" states. Fig.~\ref{results} displays the full Poincar\'{e} sphere in panel (a) for all the generated states. One notes that all the linear states live on the $S_3 = 0$ plane, as shown in (b), and all the elliptical and circular  states live on the $S_{1} = 0$ plane and off the equator (i.e., $S_{3}\neq 0$ or $\phi\neq n\pi$) as shown in panel (c).

We also examined $tr$ states intermediate between the pure {\it t} and {\it r} degrees of freedom. We prepared and measured eight different {\it tr} states of partial coherence by varying the amplitude correlation $\gamma$ in the preparation stage of our apparatus. These states are located inside the $tr$ Poincar\'{e} sphere and are illustrated in both panels (a) and (c). One notes that with the decrease of the two-path temporal coherence $\delta$ combining with the relative phase $\phi$ change, the partially coherent states are gradually rotating into the center of the Poincar\'{e} sphere, where there is a complete lack of coherence. That is, the degree of {\it tr} coherence is getting smaller and smaller. This sequence is shown explicitly in panel (c), and eventual nearness to the sphere center is quantified as $\cP_{tr} = 0.050$ in the bottom line of Table I below. \\

\noindent\textbf{Summary:} \quad We have identified the $tr$ category of optical coherence for the first time, and have described its features theoretically, and recorded those features experimentally. It is the missing member of a fundamental triad, previously completely hidden and now revealed by our derivation of the triad of coherence constraints. The members of the triad arise as in (\ref{etr})-(\ref{est}) from separate projections of the same optical field (\ref{Efield}) on its three degrees of freedom. One member is traditionally identified with the correlation of temporal amplitude with spin (ordinary polarization). Another member has only recently been identified as a hidden coherence that correlates spatial amplitude with spin. The new third member is the first optical coherence independent of spin, and arises from correlation of temporal and spatial amplitudes. Our approach establishes that there can be no more hidden optical coherences \cite{moredegrees}.

\begin{table}[b!] 
\begin{center}
\begin{tabular}{|l|l|l|l|l|l|l|l|l|}
\hline
& $\theta$ & $\phi $ &  $S_{1}$ & $S_{2}$ & $S_{3}$ &${\cal P}_{\rm tr}$
&${\cal C}_{\rm tr}$& ${\cal P}^{2}_{\rm tr}$+${\cal C}^{2}_{\rm tr}$ \\
\hline
l&$3\pi/2$ & $\ \ 0$  & $0.026$ & $-0.916$ & $\ \ 0.037$ & 0.918 &0.392&\ \ 0.996 \\
\hline
c& $\pi/2$ &$-\pi/2$ & $0.029 $ & $-0.026$ & $-0.889 $ &0.890&0.455&\ \ 0.998 \\
\hline
e&$\pi/2$ &$\pi/4$ & $0.024$ & $\ \ 0.679$ & $\ \ 0.625$ &0.923&0.384&\ \ 0.999  \\
\hline
p&$\pi/2$ &$5\pi/4$ & $0.052$ & $-0.271$ & $-0.306$  & 0.307&0.945&\ \ 0.988\\
\hline
u& $\pi/2$ &$2\pi$ & $0.042$ & $-0.025$ & $-0.013$  & 0.050&0.991&\ \ 0.985\\
\hline
\end{tabular}
\end{center}
\caption{Stokes parameters normalized to $S_0$, degree of $tr$ ``polarization" and concurrence for selected measured states. The first three are (l)inear, (c)ircular and (e)lliptical $tr$  states described by different parameters of $\theta$ and $\phi$, and the last two are (p)artial and (u)n-polarized states. The maximum standard deviation of all Stokes parameter measurements (including those not listed in the table but illustrated in Fig.~\ref{results}) is $0.042$.} \label{StokesTable}
\end{table}

The theoretical analysis leading to the discovery of the $tr$ coherence domain revealed the presence of a quantitative balance between the degree of polarization and degree of entanglement (nonseparability) of the participating vector spaces (degrees of freedom). This balance takes the form of quadratic constraints applying to all pairs of degrees of freedom previously discussed.  These constraints  $\cP^2 + \cC^2 = 1$, can be interpreted as a new expression of complementarity, and we promised experimental confirmation of the third constraint. This has been done, as recorded in the final column of Table I below, showing values for $\cP^2_{tr} + \cC^2_{tr}$ equal to 1 well within the recorded standard deviation of 0.042 for the Stokes parameters of all five rows in the Table.

Experimental entry into the domain of the third coherence required creation in the laboratory of a new form of optical coherence tomography, which we described in detail. It is of some interest to note that,  as illustrated in Fig.~\ref{setup}, we are able to prepare arbitrary states of the form (\ref{Detr}) by experimental choices of $\delta$, $\theta$ and $\phi$. Measurements of the parameters $\theta$ and $\delta$ can be realized by registering individual as well as combined intensities of P1 and P2 at the preparation stage after passing through an appropriately rotated (45 degree) Dove prism. This process is independent of the subsequent coherence tomography procedure, and gives a measurement of $\cC_{tr}$ independent of the measurement of $\cP_{tr}$, using the same modes. 

Two final comments: (a) Our discovery of the third member of the fundamental coherence triad removes the mystery of hidden coherences. They are a real consequence of the contextual character of coherence. Context matters! Coherence between a pair of degrees of freedom is isolated by projection of the other independent degree of freedom. If one of the pairs becomes accessible by an appropriate experimental projection, the others are made inaccessible, i.e., become ``hidden", even if present in the unprojected field. (b) Entirely new questions arise from the recognition of the triad of two-way coherences. We have shown that each of the two-way coherences is accompanied by a different complementarity. It is fascinating to ask whether all three degrees of freedom can be treated together, none traced or projected from consideration. This points to a completely new avenue of coherence study. We expect that our results foretell a new three-way interpretation of coherence, which will enlarge the meaning of complementarity itself. Work in this direction is under way \cite{3-way}, with results to be reported subsequently. \\

\noindent\textbf{Acknowledgement:} \quad
Support is acknowledged from a University of Rochester Research Award, ARO W911NF-14-1-063, ONR N00014-14-1-0260, as well as NSF grants PHY-1203931, PHY-1505189, and
INSPIRE PHY-1539859.


\end{document}